\documentclass[
a4paper,
11pt]{article}

\usepackage{fullpage}

\usepackage{booktabs}
\usepackage{caption}

\usepackage{mathtools}
\usepackage{color}
\usepackage{graphicx}
\usepackage{epsf}
\usepackage{graphicx,epsfig}
\usepackage{multirow}

\usepackage{amsmath, amsthm, amssymb}

\usepackage[T1]{fontenc}

\usepackage{epsfig}
\usepackage{cite}
\usepackage{color,colordvi}
\newcommand{\be}{\begin{eqnarray}}
\newcommand{\ee}{\end{eqnarray}}
\newcommand{\bi}{\begin{itemize}}
\newcommand{\ei}{\end{itemize}}

\newcounter{hran}

\makeatletter
\renewcommand\section{\@startsection {section}{1}{\z@}%
                               {-3.5ex \@plus -1ex \@minus -.2ex}%
                               {2.3ex \@plus.2ex}%
                               {\normalfont\large\bfseries}}
\makeatother

\begin{document}
\vspace{5mm}
\vspace{0.5cm}

\def\thefootnote{\fnsymbol{footnote}}


%
\begin{titlepage}
	\thispagestyle{empty}
	\begin{flushright}
		\hfill{CERN-TH-2107-158, DFPD-2017/TH/11,
	MPP-2017-161,
	LMU-ASC 43/17}
	\end{flushright}

	\vspace{30pt}
	
{\centering	
	    { \Large{\bf 
	
	 Non-linear Realizations and  Higher Curvature Supergravity
	    }}
		
		\vspace{40pt}

		F. Farakos$^{1,2}$, S. Ferrara$^{3,4,5}$, A. Kehagias$^6$ 
		and D. L\"ust$^{7,8}$

		\vspace{25pt}

			$^1${\it  Dipartimento di Fisica e Astronomia ``Galileo Galilei''\\
			Universit\`a di Padova, Via Marzolo 8, 35131 Padova, ITALY}

		\vspace{15pt}

			$^2${\it   INFN, Sezione di Padova, 
			Via Marzolo 8, 35131 Padova, ITALY}

		\vspace{15pt}

			$^3${\it Department of Theoretical Physics CH - 1211 Geneva 
              23, SWITZERLAND}\\

		\vspace{15pt}

			$^4${\it  INFN - Laboratori Nazionali di Frascati
Via Enrico Fermi 40, I-00044 Frascati, ITALY}\\

		\vspace{15pt}
		
$^5${\it Department of Physics and Astronomy, Mani L. Bhaumik Institute for Theoretical Physics U.C.L.A., Los Angeles CA 90095-1547, USA	}
   
		\vspace{15pt}	
		
$^6${\it
Physics Division, National Technical University of Athens,\\
 15780 Zografou Campus, Athens, GREECE
}\\

		\vspace{15pt}
$^7${\it 
Arnold Sommerfeld Center for Theoretical Physics\\
Theresienstra\ss e 37, 80333 M\"unchen, GERMANY}\\

		\vspace{15pt}
$^8${\it Max-Planck-Institut f\"ur Physik F\"ohringer Ring 6, 80805 
M\"unchen, GERMANY}

		\vspace{40pt}

}


\hrule \vspace{0.3cm}
 \noindent \textbf{Abstract} \\[0.3cm]
\noindent 
We focus on non-linear realizations of local supersymmetry as obtained by using constrained superfields in supergravity. 
New constraints, beyond those of rigid supersymmetry, are obtained whenever curvature multiplets are affected as well as higher derivative interactions are introduced. In particular, a new constraint, which removes a very massive gravitino is introduced, 
and in the rigid limit it merely reduces to an explicit supersymmetry breaking. Higher curvature supergravities free of ghosts and instabilities are also obtained in this way. 
 Finally, we consider direct coupling of the goldstino multiplet to the super Gauss--Bonnet multiplet and discuss the emergence of a new scalar degree of freedom. 
\vspace{0.5cm}  \hrule
\vskip 1cm

\bigskip

\end{titlepage}


\def\thefootnote{\arabic{footnote}}
\setcounter{footnote}{0}

\def\ls{\left[}
\def\rs{\right]}
\def\lc{\left\{}
\def\rc{\right\}}

\def\p{\partial}

\def\S{\Sigma}

\def\s{\sigma}

\def\O{\Omega}

\def\a{\alpha}
\def\b{\beta}
\def\g{\gamma}

\def\ad{{\dot \alpha}}
\def\bd{{\dot \beta}}
\def\gd{{\dot \gamma}}

\def\nn{\nonumber}

\baselineskip 6 mm


\newpage

\tableofcontents



\section{Introduction}

Higher curvature supergravity has been revived recently, especially 
motivated from cosmology \cite{R22,staro,TNi,Cecotti1,Cecotti2}. 
In particular, $R+R^2$ theories have been extensively studied as they propagate besides the spin-2 graviton, also a massive scalar mode \cite{Stelle,FGN}. 
This scalar, often referred to as scalaron, is identified with the inflaton in Starobinsky inflation \cite{Starobinsky:1980te}. 
Other higher curvature actions, like Weyl square, might give rise to ghost states. 
For example the latter introduces a massive spin-2 ghost in the spectrum, and therefore is problematic. 
The embedding of the various cosmological models within a UV complete framework is 
eventually essential, 
and might also lead to a deeper understanding of the effective theory of inflation \cite{Lyth:1998xn}. 
Four-dimensional N=1 supergravity theories are particularly motivated for such a study, 
because they arise as the low energy limits of string theory \cite{Blumenhagen:2013fgp}. 
Once supergravity is treated as an effective theory, 
then the standard matter-coupled action \cite{CFGvP,Wess:1992cp,Freedman:2012zz} 
will generically receive higher derivative corrections coming from the UV physics \cite{Cecotti1,Cecotti2}. 
Therefore their consistent embedding into 4D N=1 supergravity deserves a careful investigation.

When attempting to ``supersymmetrize'' higher derivative gravity coupled to matter, 
even only the consistent interactions, on top of the modification of the scalar potential \cite{CFG,EMP,CLW,KY}, 
one is faced with an unavoidable problem: 
the new terms added to the action often give rise to ghosts \cite{Cecotti1,Cecotti2,Baumann:2011nm}. While this phenomenon does not happen in supersymmetric vacua \cite{Cecotti1}, it manifests when supersymmetry is broken which is always the case when it is non-linearly realized. 
The ghosts arise from unwanted kinetic mixings of the scalars to gravity, 
as off-diagonal kinetic terms, 
or as auxiliary fields which become propagating. 
Even though there do exist islands of consistent supersymmetric 
higher derivative theories, 
one would like to have at hand a tool, which will allow to study the effective description 
of all the possible higher derivative interactions within supergravity. 
Essentially, one needs a way to eliminate the ghost states from the spectrum, 
which is the basic motivation of the present study. 
In particular we will focus on the couplings of the inflaton to quadratic curvature of the form 
\be
\label{intro1}
a_1 (\phi) \, R^2_{GB} 
+ a_2 (\phi) \, R^2_{HP} 
+ a_3 (\phi) \, C^2 \, , 
\ee
where $R^2_{GB}$ is the Gauss-Bonnet combination, 
$R^2_{HP}$ is the Hirzebruch--Pontryagin combination and 
$C^2$ is the square of the Weyl tensor. 
The coupling with $R^2$ has been extensively studied in the literature, 
and therefore we will not study it in detail, 
but we refer the  interested reader to \cite{staro} and to \cite{morequad}.

In effective supergravity theories where supersymmetry is spontaneously broken 
and non-linearly realized \cite{Samuel:1982uh,DSS} 
it is known how to eliminate any unwanted state from the spectrum \cite{Ferrara:2016een,Dall'Agata:2016yof}. 
Indeed, 
if the goldstino resides inside the nilpotent chiral superfield $X$, 
which satisfies $X^2 = 0$, 
one can eliminate the lowest component field of a generic superfield $Q$, 
by imposing the constraint $X \overline X \, Q = 0$. 
Constrained superfields have been already extensively studied for the effective description of 
supersymmetric theories during inflation \cite{nilinfl}. 
This use is justified, 
because during inflation various component fields will become heavy and decouple from the inflationary sector, 
and a way to describe this decoupling is by using constrained superfields. 
Drawing inspiration from this property of constrained superfields, 
we impose superspace constraints 
in such a way that the unwanted ghost fields in supersymmetric theories 
with higher derivatives are eliminated from the spectrum. 
In particular, we focus on the superspace Lagrangians which contain 
known consistent terms in their bosonic sector, 
but due to the supersymmetric completion they contain ghosts as well. 
We show that the ghosts are eliminated from the spectrum due to the superfield constraints. 
These constraints are imposed both on the matter multiplets and on the supergravity multiplet. 
The constraints on the supergravity multiplet are imposed 
to eliminate the supergravity auxiliary fields 
which would otherwise give rise to ghosts \cite{Cecotti1,Cecotti2}.

Another aspect of our work is related to the gravitino. 
First, we will construct a new term which generates a pure gravitino mass, 
and in the global limit it will generate higher order interaction terms for the goldstino. 
Second, we introduce a new constraint on the nilpotent $X$ superfield 
which eliminates the gravitino from the low energy supergravity theory. 
In the global limit this constraint will essentially eliminate the goldstino 
therefore will lead to a stronger constraint on $X$ which will change it into a {\it spurion}.  In other words, the constraint  we are presenting here
have no analogue in rigid susy. In the absence of the gravitino our constraint  makes the goldstino to vanish
so only the spurion vev would remain.

This work is organized as follows. In section 2, the basic ingredients of local superspace geometry and the notation are introduced together with non-linear constraints on matter and curvature multiplets. In particular, a new constraint on the Volkov-Akulov multiplet decouples the gravitino. 
In section 3 couplings of linear and non-linear chiral multiplets to higher curvature invariants are introduced. 
An extreme case is the  Volkov-Akulov Lagrangian augmented by its coupling to the Gauss-Bonnet super invariant. This theory implies the existence of a new scalar degree of freedom, dual to a specific $f(R)$ gravity encoded as part of the supersymmetrized Gauss-Bonnet invariant.  The paper ends with some concluding remarks in section 4.

\section{Supergravity, matter, and constrained superfields}

\subsection{Curvature multiplets, matter multiplets and invariant actions}

In this work we use the superspace formalism of old-minimal supergravity
\cite{auxiliary}. Our conventions can be found in \cite{Wess:1992cp}, 
whereas the complete old-minimal supergravity curvature multiplets can be found in  \cite{Ferrara:1988qx}. 
The construction of supergravity relies on two superspace fields, 
the super-vielbein $E_M^{\ \, A}$ and the super-connection $\phi_{M A}^{\ \ \ \, B}$. 
The independent and propagating degrees of freedom of the supergravity multiplet are
the vielbein $e_m^{\ a}$ and the gravitino $\psi_m^{\ \alpha}$. 
From the super-vielbein and the super-connection we can built the torsion tensor superfield ${\cal T}_{NM}^{\ \ \, A}$ 
and the curvature tensor superfield ${\cal R}_{NMA}^{\ \ \ \ \ \, B}$ \cite{GWZ}.

The components of the torsion  and the curvature  superfields are not independent, 
and they can be expressed in terms of three independent superfields \cite{WZ1}. 
These superfields are 
the Ricci superfield  ${\cal R}$, 
the $B_{\alpha \dot \alpha}$ superfield, 
and the Weyl superfield ${\cal W}_{\alpha \beta \gamma}$ and their properties
 can be found in \cite{Wess:1992cp,Cecotti1,Ferrara:1988qx}. 
This reduction is achieved by employing the torsion constraints 
and the Bianchi identities for the torsion and the curvature. 
Then the solution of the Bianchi identities under the torsion constraints, 
not only leads to the three aforementioned curvature superfields, 
but also gives conditions on these superfields among themselves. 
From the Bianchi identities (the full list can be found in \cite{Wess:1992cp}), 
one can relate the higher components of the irreducible curvature superfields to the 
fields of the supergravity multiplet. 
These components are particularly important for our discussion because they contain various curvature tensors.  
The independent  superfields of the supergravity multiplet are: 
\begin{enumerate}
\item
{\it The Ricci superfield  ${\cal R}$}.   It is a chiral superfield
\be
\label{RG1}
\overline{ {\cal D}}_{\dot{\alpha}} {\cal R} = 0 \ , 
\ee
and in the lowest component it contains the scalar auxiliary field of the supergravity multiplet 
\be
{\cal R} | = -\frac{1}{6} M \, .  
\ee
The  highest component of the ${\cal R}$ superfield has bosonic sector 
\be
{\cal D}^2 {\cal R} | =  -\frac{1}{3} R + \frac{4}{9} M \overline M + \frac{2}{9} b^a b_a 
-\frac{2 i}{3} e_{a}^{\ m} {\cal D}_m b^a . 
\ee
Here the abbreviation ${\cal O}|$ stands for ${\cal O} |_{\theta=\overline \theta = 0}$, 
which means to set the Grassmann parameters to vanish. 

\item 
{\it The $B_{\alpha \dot \alpha}$ superfield}.  It has the properties 
\be
\label{RG2}
( B_{\alpha \dot \alpha})^* =   B_{\alpha \dot \alpha} \, , \  \ 
\overline {\cal D} ^{\dot{\alpha}}  B_{\alpha \dot \alpha} =  {\cal D}_{ \alpha} {\cal R} \, ,
\ee
and its lowest component  is the second auxiliary field of old-minimal supergravity 
\be
B_a | = -\frac13 b_a \, , 
\ee
where $b_a$ is a real vector. 
The higher components of the superfield $ B_a$ have bosonic sectors 
\be
\begin{split}
{\cal D}^2 B_a | &= -\frac{2 i}{3} {\cal D}_a \overline M + \frac23 b_a \overline M \, , 
\\
\overline \s^{\dot \alpha \alpha}_b {\cal D}_\alpha \overline {\cal D}_{\dot \alpha}    B_a | &= 
\left( \frac16 R +\frac19 M \overline M + \frac19 b^c b_c \right) \eta_{a b} 
- R_{ab} 
-\frac{2 i}{3} {\cal D}_b b_a + \frac29 b_a b_b  - \frac13 \epsilon^{cd}_{\ \ ab} {\cal D}_c b_d \, , 
\\
{\cal D}^2 \overline {\cal D}^2  B_a | &= 
-4 i  {\cal D}_a  {\cal D}^2 {\cal R} |   
+ 4   B_a| \, {\cal D}^2 {\cal R} | 
+ 16 i \overline {\cal R}| \, {\cal D}_a {\cal R}| 
+  12 {\cal R} | \, {\cal D}^2  B_a |  \, . 
\end{split}
\ee

\item 
{\it The Weyl superfield ${\cal W}_{\alpha \beta \gamma}$}. It satisfies
\be
\label{WRG1}
{\cal W}_{\alpha \beta \gamma} = {\cal W}_{(\alpha \beta \gamma)} \  , \  
({\cal W}_{\alpha \beta \gamma})^* = \overline {\cal W}_{\dot \alpha \dot \beta \dot \gamma} \  , \   
\overline{ {\cal D}}_{\dot{\delta}} {\cal W}_{\alpha \beta \gamma}= 0  \ , 
\ee
and there is an algebraic identity which relates it to the other two superfields 
\be
\label{WRG2}
{\cal D}^\alpha {\cal W}_{\alpha \beta \gamma} 
+ \frac{i}{2} \left( \epsilon^{\dot \beta \dot \gamma} {\cal D}_{\beta \dot \beta}   B_{\gamma \dot \gamma} 
+ \epsilon^{\dot \beta \dot \gamma} {\cal D}_{\gamma \dot \beta}   B_{\beta \dot \gamma}  \right)  = 0  \, . 
\ee 
This superfield  has the following pure bosonic contributions 
\be
{\cal D}_{\gamma} {\cal W}_{\delta \epsilon \alpha} |=  
\frac16 \left( \frac{i}{4}  \epsilon_{\gamma \delta} {\cal D}_{\epsilon \dot \epsilon} b_{\alpha}^{\dot \epsilon} 
-\frac{1}{4} \sigma^{de \ \beta}_{\ \delta} \epsilon_{\beta \epsilon} 
\sigma^{ba \ \rho}_{\ \alpha} \epsilon_{\rho \gamma} R_{deba}  \right) + (\delta \, \epsilon \, \alpha) \, ,  
\ee
where $R_{abcd} =e_{a}^{\ m} e_{b}^{\ n}  R_{mncd}$ and $(\delta \, \epsilon \, \alpha)$ 
refers to the symmetrization with respect to the fermionic 
indices $\delta$, $\epsilon$ and $\alpha$ and contains five  terms.

\end{enumerate}

\noindent
The previous formulae were found at the linearized level in \cite{Ferrara:1977mv} and in the superconformal tensor calculus in \cite{FTSvP}. 

We now turn to the description of {\it scalar matter}. 
A generic chiral superfield $\Phi$ is defined by 
$\overline {\cal D}_{\dot \alpha} \Phi = 0$,  
and the component fields it contains can be defined by projection as  
\be 
\Phi|  = A  \, , \ \ 
\frac{1}{\sqrt{2}} {\cal D}_\alpha \Phi|  = \chi_\alpha \, , \ \ 
- \frac14 {\cal D}^2 \Phi|  = F \, . 
\ee
Here $A$ is a complex scalar, 
$\chi$ is a fermion and $F$ is the matter multiplet auxiliary field. 
Using the new $\Theta$ variables expansion we can also write the chiral superfield $\Phi$ as 
\be
\Phi = A + \sqrt{2} \, \Theta^\alpha \chi_\alpha + \Theta^2 F . 
\ee
We will often use the notation $\psi^2 = \psi \psi = \psi^\alpha \psi_\alpha$ 
for fermionic objects. 
In this work we will use only chiral superfields to describe the matter sector.

From any unconstrained Lorentz scalar complex superfield $U$ we can always construct a chiral superfield $\Xi$, 
with the use of the chiral projection, 
as 
\be
\Xi  = -\frac14  (  \overline {\cal D}^2 - 8 {\cal R}) U .  
\ee 
This chiral superfield has component expansion 
\be
\Xi = A^\Xi + \sqrt{2} \, \Theta^\alpha \chi^\Xi_\alpha + \Theta^2 F^\Xi \, . 
\ee
To built actions invariant under the local supersymmetry, 
one has to employ density superfields. 
In particular, 
we will use here the superspace field $2 {\cal{E}}$ which is a chiral density, 
and is defined as  
\be
\label{2Ep}
2{\cal{E}}=e \left\{ 1+ i\Theta \sigma^{a} \bar{\psi}_a 
-\Theta \Theta \Big{(} \overline M +\bar{\psi}_a\bar{\sigma}^{ab}\bar{\psi}_b 
\Big{)} \right\} . 
\ee
The property of the chiral density is that, 
when multiplied with chiral superfields, 
the product is again a chiral density. 
Then the superspace Lagrangian 
\be
\label{Lxi}
{\cal L} =   \int d^2 \Theta \, 2 {\cal{E}} \, \Xi \, + c.c.  
=  \int d^2 \Theta \, 2 {\cal{E}} \ls -\frac14  (  \overline {\cal D}^2 - 8 {\cal R}) \, U \rs + c.c. 
\ee
will be invariant under local supersymmetry up to boundary terms. 
In component form the Lagrangian \eqref{Lxi} 
will read  
\be
\label{helper}
e^{-1} {\cal L} = F^\Xi 
- \frac{i}{\sqrt 2} \, \chi^\Xi \sigma^a \overline \psi_a 
- \Big{(} 
\overline M +\bar{\psi}_a\bar{\sigma}^{ab}\bar{\psi}_b 
\Big{)} A^\Xi 
+ c.c.  
\ee
To find the free supergravity for example one has to set $\Xi = -3 {\cal R}$, 
and for the superspace cosmological constant we have $\Xi=W_0$, where $W_0$ is a complex constant.

The matter sector chiral superfields are denoted collectively  as $\Phi^I$ for $I=1,\cdots n$.  
The generic coupling (up to two derivatives) of a chiral model is 
\be
\label{chiralmodel}
{\cal L}_0 =  \int d^2 \Theta \, 2 {\cal E} \, 
\ls -\frac18 (\overline {\cal D}^2 - 8 {\cal R}) \O(\Phi^I, \overline \Phi^{\overline J} ) + W(\Phi^I) \rs + c.c. \, , 
\ee
where $\O(\Phi^I, \overline \Phi^{\overline J} )$ is a hermitian function and $W(\Phi^I)$ is a holomorphic function of the chiral superfields. 
The bosonic sector of (\ref{chiralmodel}) is 
\be
\begin{split}
\label{compcm}
e^{-1} {\cal L}_0 = & \,  \frac16 \O  R -\frac19 \O  b^a b_a 
- \Omega_{IJ} \p_m A^I \p^m \overline A^{\overline J}
-\frac{i}{3} (\O_I \p_m A^I -  \O_{\overline J} \p_m \overline A^{\overline J} ) b^m 
\\
&+ \frac19 \O M \overline M -\frac13 \O_{\overline J} \overline F^{\overline J}  \overline M - \frac13 \O_I F^I M 
+ \O_{I \overline J} F^I \overline F^{\overline J} 
\\
& - W \overline M - \overline W M + W_I F^I + \overline W_{\overline J} \overline F^{\overline J}  \, , 
\end{split}
\ee 
where ${\cal O}_J = \frac{\p {\cal O}}{\p A^J}$ and so on. 
Once we integrate out the auxiliary fields and rescale the metric 
by $g_{mn} \rightarrow -\frac{3}{\Omega} g_{mn}$ the bosonic sector gets the familiar form 
which can be found in \cite{Wess:1992cp}, 
where the K\"ahler potential is given by 
\be
K = -3 \, \text{log}\left(-\frac13 \Omega \right)  . 
\ee
In our discussion though, 
it is essential to have the matter sector in the original form given by \eqref{compcm}.  
This is because there will be auxiliary fields which cannot be integrated out, 
but one has to eliminate then by superspace constraints.

\subsection{Non-linear realizations}

In a supergravity setup, 
it is known that when supersymmetry is spontaneously broken 
and the sgoldstino is integrated out\footnote{In global supersymmetry for F-term and/or D-term breaking it has been 
shown in \cite{nodecoupling} that a chiral superfield with $X^2=0$ always appears at low energy.}, 
one can effectively describe the supersymmetry breaking sector by the use of the 
constrained chiral superfield $X$, 
which satisfies \cite{X2} 
\be
\label{X2}
X^2 =0 . 
\ee
This superfield has the $\Theta$ expansion 
\be
\label{X}
X = \frac{{G} {G}}{2 F^X} + \sqrt{2} \, \Theta^\alpha {G}_\alpha + \Theta \Theta F^X   . 
\ee 
The goldstino is the fermion ${G}_\alpha$ and to have a consistent non-linear realization it has to satisfy that 
\be
\label{condf}
\langle F^X \rangle \ne 0 . 
\ee
The minimal model which can describe this theory 
is constructed by a flat K\"ahler potential and a linear superpotential 
in $X$ \cite{DSS}. 
Once we have a chiral superfield $X$ with these properties we can eliminate various 
component fields from other multiplets. 
These multiplets can be curvature \cite{Cribiori:2016qif} or they can be matter \cite{Ferrara:2016een,Dall'Agata:2016yof}. 
The underlying principle is that the irreducible constraint 
\be
X \overline X \, Q = 0 , 
\ee
eliminates \emph{only} the lowest component of the superfield $Q$ from the spectrum 
without imposing any constraint on the higher component fields \cite{Dall'Agata:2016yof}. 
It has been also illustrated in \cite{Cribiori:2016qif} that these constraints are equivalent to eliminating 
the specific component field by using the CCWZ procedure, 
therefore providing a non-trivial cross-check for the self-consistency.

In this work we will employ a very specific constrained matter chiral superfield ${\cal A}$ 
which will contain the real scalar $\phi$.  
In fact this constrained superfield has found its way in many applications of inflation in supergravity 
\cite{calAsuperfield}, 
and together with $X$  they provide the minimal setup for our study. 
It can be written as 
\be
{\cal A} = \phi + i b + \sqrt 2 \Theta \chi^{\cal A} + 
\Theta^2 F^{\cal A} \, , \label{AA}
\ee
and we impose the constraint  \cite{Komargodski:2009rz} 
\be
\label{Ac}
X \left( {\cal A} - \overline {\cal A} \right) = 0 . 
\ee
As a consequence of  (\ref{Ac}),   
the imaginary part of the lowest component of ${\cal A}$, the auxiliary field $F^{\cal A}$, 
and the fermionic component field $\chi^{\cal A}_\alpha$ are removed from the spectrum. 
The most convenient way to verify this is to realize that from \eqref{Ac} we can produce the 
irreducible constraints 
\be
X \overline X  \left( {\cal A} - \overline {\cal A} \right) = 0 \ , \  
X \overline X \,  {\cal D}_\alpha  {\cal A}  = 0 \ , \ 
X \overline X \,  {\cal D}^2  {\cal A}  = 0 \, . 
\ee
To summarize, the superfield ${\cal A}$ contains only a real scalar degree of freedom $\phi={\rm Re}{\cal A}| $ while the  
other fields of the ${\cal A}$ multiplet are solved in terms of the goldstino.

As we said, the supergravity multiplet also contains auxiliary degrees of freedom which for the old-minimal formulation are the 
complex scalar $M$ and the real vector $b_m$. 
One can also impose constraints on the multiplets which contain these fields, and thus eliminating them from the spectrum. 
In particular we may impose  
\be
\label{XXG}
X \overline X \,  B_a = 0 , 
\ee
which implies for the supergravity auxiliary field $b_a = {\cal O}({G}, \overline{G})$. 
Indeed, following \cite{Cribiori:2016qif} we see that  \eqref{XXG} gives the superspace equation 
\begin{equation}
\label{XXGGG}
B_a = -2 \frac{\overline{\cal D}_{\dot\alpha}\overline X}{\overline{\cal D}^2 \overline X}\overline{\cal D}^{\dot \alpha}B_a - \overline X \frac{\overline{\cal D}^2 B_a}{\overline{\cal D}^2\overline X}-2\frac{{\cal D^\alpha}X}{{\cal D}^2 X\,\overline{\cal D}^2\overline X} {\cal D}_\alpha \overline{\cal D}^2(\overline X B_a) -\frac{X}{{\cal D}^2 X \overline{\cal D}^2 \overline X}{\cal D}^2 \overline{\cal D}^2 (\overline X B_a). 
\end{equation}
This equation will give an expression for $b_a$ which has to be solved iteratively. 
Notice that by expanding the right hand side of \eqref{XXGGG} we find 
that it contains terms of the form 
\be 
\label{bbbb}
b_a \supset 
+ \frac{\sqrt 2 G^\alpha}{2 F} ({\cal D}_\alpha B_a | ) 
-\frac12 \frac{G^\alpha \overline G^{\dot \alpha}}{F \overline F} 
\left( {\cal D}_\alpha \overline{\cal D}_{\dot \alpha} B_a| \right) \, , 
\ee 
once we turn to its component field expansion. 
Moreover due to the structure of (\ref{XXG}) no other component field, except the lowest one, of the $B_a$ 
superfield is constrained. 
Indeed, as illustrated in \cite{Dall'Agata:2016yof} for the constraints of the form $|X|^2 Q=0$, 
all the possible projections with any combination of ${\cal D}_\alpha$ or $\overline{\cal D}_{\dot \alpha}$ will serve  
only as consistency conditions {\it except} when acting with the maximum projection ${\cal D}^2 \overline{\cal D}^2$ 
which gives a single constraint on $Q|$. 
For example if we act with ${\cal D}_\alpha$  on \eqref{XXG} 
and projecting to components we find 
\be
\label{XXGbb}
(\sqrt 2 G_\alpha) \frac{\overline G^2}{2 \overline F} \,  b_a 
+ \frac{G^2 \overline G^2}{4 F \overline F}  \left( {\cal D}_\alpha B_a | \right) = 0 \, ,  
\ee 
which is just a consistency condition for the full solution of $b_a$ and \emph{not} 
a constraint on ${\cal D}_\alpha B_a|$, 
as can be checked by replacing \eqref{bbbb} into \eqref{XXGbb}. 
Note finally that since the component ${\cal D}_\alpha B_a|$ is related to the gravitino field strength,  it means that
if it had been removed from the spectrum it would be a sign of inconsistency. 
In addition one can have 
\be
\label{XRc}
X \left( {\cal R} + \frac{c}{6} \right) = 0 ,
\ee
where $c$ is a complex constant, which gives $M = c + {\cal O}({G})$. 
These constraints have been studied extensively in \cite{Cribiori:2016qif}.

In closing this subsection, 
let us see how the non-linear realization of supersymmetry allows the construction of 
a vast number of new terms.\footnote{The relation between the formalism with the nilpotent $X$ 
and the spinor superfield of \cite{Samuel:1982uh} (which allows to build a vast number of new terms) 
has been thoroughly explained in \cite{Cribiori:2016hdz,Bandos:2016xyu}.} 
Consider a superspace Lagrangian of the form 
\be
{\cal L}_{\cal Z} = 16 \int d^4 \theta \, E \,  \frac{X}{{\cal D}^2 X} 
\frac{\overline X}{\overline{\cal D}^2 \overline X}  \, 
{\cal Z} \left(\Phi^I \, , {\cal D}_\alpha \Phi^I
\, , {\cal D}^2 \Phi^I 
\, ,  {\cal D}^2 X \, ,  \cdots \right) \, . 
\ee 
Then in the $G=0$ gauge we will have 
\be
e^{-1} {\cal L}_{\cal Z} = {\cal Z} \left( A^I \, , \sqrt 2 \chi_\alpha^I
\, , -4 F^I 
\, ,  -4 F^X \, , \cdots  \right) \, . 
\ee
Such a possibility clearly exists due to the fact that all the goldstino 
can do is to work as a compensator which can always be used to {\it dress} any component field 
and to construct a supersymmetric Lagrangian of generic form. 
The simplest Lagrangian one can construct is by setting ${\cal Z} = - \Lambda$, 
where $\Lambda$ is a real constant, 
which would simply give \cite{Samuel:1982uh} 
\be
{\cal L}_\Lambda = - 16 \Lambda \, \int d^4 \theta \, E \,  \frac{X}{{\cal D}^2 X} 
\frac{\overline X}{\overline{\cal D}^2 \overline X} \, . 
\ee 
The component form expansion is 
\be
\label{LLcc}
{\cal L}_\Lambda = - e \Lambda   
- i e \Lambda 
\left( \frac{G^\alpha}{F} \right) 
\s^m_{\alpha \dot \alpha} {\cal D}_m 
\left( \frac{\overline G_{\dot \alpha}}{\overline F} \right) 
+ \cdots  
\ee
which in the $G=0$ gauge will provide only a contribution to the vacuum energy. 
Notice that the same constant $\Lambda$ which changes the vacuum energy in \eqref{LLcc} multiplies 
the kinetic term of the goldstino and therefore will change the supersymmetry breaking scale 
once the goldstino is canonically normalized \cite{CFG}. 
In the next subsection we will use this method to construct a pure gravitino mass and study its properties, 
and we will see how the supersymmetry breaking scale is generically altered.

\subsection{Gravitino decoupling}

Until now we have discussed how to eliminate the bosonic auxiliary fields of the supergravity multiplet. 
We now discuss the decoupling of the gravitino. 
As we will see to remove the gravitino from the theory one has to impose a constraint on the 
nilpotent chiral superfield $X$.

Let us first assume that supersymmetry  is broken from the nilpotent chiral superfield $X$, 
which is the only chiral multiplet in the theory that is coupled to supergravity. 
Then we can generically include in our effective Lagrangian a higher derivative term 
\be
\label{gmass1}
{\cal L}_{m_{3/2}} = (16)^2 \, \zeta \int d^4 \theta \, E \, 
\left[  
\frac{X}{{\cal D}^2 X} 
\frac{\overline X}{\overline{\cal D}^2 \overline X} 
\, 
{\cal D}_a \left( \frac{{\cal D}^\alpha X}{{\cal D}^2 X} \right) 
\s^{ab \ \beta}_{\ \ \alpha} \, 
{\cal D}_b \left( \frac{{\cal D}_\beta X}{{\cal D}^2 X} \right)  
+ c. c. 
\right] 
\, , 
\ee
where $\zeta$ is a real constant parameter of mass dimension three: $[\zeta]=3$. 
In component form we find that 
\be
\label{gmass2}
e^{-1} {\cal L}_{m_{3/2}} = \frac{\zeta}{ M_P^2} \, \psi_a \s^{ab}  \psi_b 
+ \frac{\zeta}{ M_P^2}   \, \overline \psi_a \overline \s^{ab}  \overline \psi_b + {\cal O}(G,\overline G) \, , 
\ee
where we have restored the Planck scale $M_P$.
One can easily verify that \eqref{gmass1} gives \eqref{gmass2} by taking into account that 
\be
{\cal D}_a \left( \frac{{\cal D}^\alpha X}{{\cal D}^2 X} \right)  
= 
\frac{1}{4 M_P} \psi_a^{\, \alpha} 
- \frac{1}{2 \sqrt 2} \, e_{a}^{\, m} D_m \left( \frac{G^\alpha}{ F^X} \right) + \cdots  \, . 
\ee
This term now introduces a non-supersymmetric mass term to the gravitino. 
Independent gravitino mass terms giving \eqref{gmass2} were also constructed in \cite{Bandos:2016xyu} using different methods. 
Notice that in the global limit the superspace Lagrangian \eqref{gmass1}  
becomes 
\be
\label{gmass3}
{\cal L}_{m_{3/2}} = (16)^2 \, \zeta \int d^4 \theta \,  
\left[  
\frac{X}{D^2 X} 
\frac{\overline X}{\overline D^2 \overline X} 
\, 
\partial_m \left( \frac{D^\alpha X}{D^2 X} \right) 
\s^{mn \ \beta}_{\ \ \alpha} \, 
\partial_n \left( \frac{D_\beta X}{D^2 X} \right)  
+ c. c. 
\right] 
\, , 
\ee
therefore it will contribute to higher order goldstino interactions. 
If for example we have a supergravity coupled to $X$ of the standard form  \eqref{chiralmodel} 
with 
\be
\label{KW1}
K = X \overline X \ , \ W = f X + W_0 \, , 
\ee
and the additional term \eqref{gmass1},  
 the component form of the total Lagrangian turns out to be (in the $G=0$ gauge) 
\begin{equation}
\label{LL11}
\begin{split}
e^{-1} {\cal L} = &  -\frac12 M_P^2 R 
+ \frac{1}{2} \epsilon^{klmn} (\overline \psi_k \overline \sigma_l {\cal D}_m \psi_n - \psi_k \sigma_l {\cal D}_m \overline \psi_n) 
\\[2mm]
&  
- \left( \frac{W_0 - \overline \zeta}{M_P^2}  \right) \, \overline \psi_a \overline \sigma^{ab} \overline \psi_b 
- \left( \frac{\overline{W_0} - \zeta}{M_P^2}  \right) \, \psi_a  \sigma^{ab}  \psi_b 
- |f|^2 + 3 \frac{ |W_0|^2}{M_P^2}   \, . 
\end{split}
\end{equation}
Notice that the mass dimensions are: $[f]=2$ and $[W_0]=3$.

We set $M_P=1$ fron now on. 
In the Lagrangian \eqref{LL11} 
one can easily integrate out the gravitino by taking 
the parameter $\zeta$ to be large. 
In this case the mass term dominates over the momenta and in this limit 
$\psi_m$ appears algebraically and can be integrated out. 
The constraint we need to impose in order to describe this decoupling is 
\be
\label{XXgrav}
X \overline X \, {\cal D}_b \left( \frac{{\cal D}_\beta X}{{\cal D}^2 X} \right) = 0 \, . 
\ee
The consistency of the constraint \eqref{XXgrav}  is guaranteed by the fact that the lowest component 
of ${\cal D}_a \left( {\cal D}_\beta X / {\cal D}^2 X \right)$ 
starts with the component field $\psi_m^\alpha$ and not with derivatives. 
By studying the highest component field of  \eqref{XXgrav} 
one finds the full constraint, which then has to be solved iteratively, and is given by 
\begin{equation}
\begin{aligned}
{\cal D}_b \left( \frac{{\cal D}_\beta X}{{\cal D}^2 X} \right)  = & -2 \frac{\overline{\cal D}_{\dot\alpha}\overline X}{\overline{\cal D}^2 \overline X} 
\, \overline{\cal D}^{\dot \alpha} {\cal D}_b \left( \frac{{\cal D}_\beta X}{{\cal D}^2 X} \right)  
- \frac{\overline X }{\overline{\cal D}^2\overline X}  \, \overline{\cal D}^2 {\cal D}_b \left( \frac{{\cal D}_\beta X}{{\cal D}^2 X} \right)
\\
&- 2\frac{{\cal D^\alpha}X}{{\cal D}^2 X\,\overline{\cal D}^2\overline X} {\cal D}_\alpha \overline{\cal D}^2 \left[ \overline X \, 
{\cal D}_b \left( \frac{{\cal D}_\beta X}{{\cal D}^2 X} \right) \right] 
\\
& - \frac{X}{{\cal D}^2 X \overline{\cal D}^2 \overline X}{\cal D}^2 \overline{\cal D}^2 
\left[ 
\overline X \, 
{\cal D}_b \left( \frac{{\cal D}_\beta X}{{\cal D}^2 X} \right) 
\right] .
\end{aligned}
\end{equation} 
To illustrate the properties of the constraint \eqref{XXgrav}, 
the lowest component is 
\be
G^2 \overline G^2 \left( \psi_a^{\, \alpha} - \sqrt 2 \, e_{a}^{\, m} \partial_m  G^\alpha / F^X \right) =0 \, ,  
\ee
which means that once the gravitino will be eliminated in this way, 
it will have the form 
\be
\psi_a^{\, \alpha} = \sqrt 2 \, {\cal D}_a \left( \frac{G^\alpha}{ F^X} \right) + \cdots 
\ee
where the dots stand for higher order goldstino terms. 
In the rest of the work we will therefore not discuss the gravitino, 
because it may be always eliminated from the spectrum in this way.

Notice that \eqref{XXgrav} is a constraint on the $X$ superfield, 
nevertheless its effect is to eliminate the gravitino from the supergravity multiplet. 
Moreover in the global limit $M_P \rightarrow \infty$, Eq.\eqref{XXgrav}  leads to an {\it explicit} supersymmetry breaking 
because it will imply $G_\alpha =0$, 
and therefore $X$ will be a spurion superfield 
\be
X = \theta^2 \langle F^X \rangle \, . 
\ee

We see that if supersymmetry is spontaneously broken, 
one can construct a gravitational multiplet which will contain only the metric $g_{mn}$ (or the vielbein). 
The other component fields which are needed for the closure of the algebra will be replaced with 
composite operators of the metric and the goldstino. 
From the properties of the supersymmetry transformations one can still go to a gauge where $G=0$. 
Interestingly, in the gauge $G=0$, 
a supergravity theory subject to the constraints \eqref{XXG}, 
\eqref{XRc}  and \eqref{XXgrav}, 
will always reduce to the  Einstein gravity with a  cosmological constant. Indeed, following \cite{Cribiori:2016qif}, for 
\begin{eqnarray}
K=X\overline X, ~~~~W=f X+m_{3/2},
\end{eqnarray}
the Lagrangian  
\be
{\cal L}_\text{E--H} =  \int d^2 \Theta \, 2 {\cal E} \left\{\, {\cal R}+
X\left[-\frac{1}{8}\left(\overline {\cal D}^2-8{\cal R}\right)
\overline X\right]+W+c.c.\right\} 
\ee
reduces to 
\begin{eqnarray}
{\cal L}_\text{H--E} =-\frac{1}{2}e\, R - e \, \Lambda \, , \label{sf1}
\end{eqnarray}
where 
\begin{eqnarray}
\Lambda=\frac{1}{3}|c|^2+|f|^2+m_{3/2}\overline c+\overline m_{3/2} c. \label{sf2}
\end{eqnarray}

Let us also discuss the supersymmetry breaking scale in a model with 
K\"ahler potential and superpotential as in \eqref{KW1}, 
and Lagrangian which contains also an explicit gravitino mass of the form \eqref{gmass1}. 
The Lagrangian \eqref{gmass1} is a higher derivative term, 
and therefore it will generically violate the standard supergravity formulas. 
Indeed, even though the K\"ahler potential in \eqref{KW1} is flat, 
the relation between the gravitino mass, 
the vacuum energy and the vacuum expectation value of the auxiliary field $F^X$ 
do \emph{not} have the traditional form since  one can see that 
\be
V \ne |F^X|^2 - 3 m_{3/2}^2 \, . 
\ee
This happens because the term \eqref{gmass1} generates 
an unusual  goldstino-gravitino mixing. 
We have 
\be
\begin{aligned}
e^{-1} {\cal L}_{m_{3/2}} |_\text{kin.mix.}  
=  4 \zeta
 \Big{[} & \hat D_a \left( \frac{G^\alpha}{\sqrt{2} F^X} \right)  
\s^{ab \ \beta}_{\ \ \alpha}  
\hat D_b \left( \frac{G_\beta}{\sqrt{2} F^X} \right) 
\\
& + \frac{i}{3} M 
\frac{\overline G_{\dot \epsilon}}{\sqrt{2} \overline F^X} 
\overline \s^{\dot \epsilon \alpha}_{a} 
\s^{ab \ \beta}_{\ \ \alpha} 
\hat D_b \left( \frac{G_\beta}{\sqrt{2} F^X} \right)
\Big{]} + c.c. \, , 
\end{aligned}
\ee
where 
\be
\hat D_a \left( \frac{G^\alpha}{\sqrt{2} F^X} \right) = 
e_a^m \left[ {\cal D}_m \left( \frac{G^\alpha}{\sqrt{2} F^X} \right) 
- \frac12 \psi_m^\alpha  \right] \, . 
\ee
These new terms are responsible for an unconventional goldstino-gravitino mixing 
which alters the standard supergravity results and therefore 
the vacuum expectation value of $F^X$ seizes to be the supersymmetry breaking scale.

To correctly identify the supersymmetry breaking scale, we define a new scale $f^S$ as 
\be
\label{FFSS} 
f^S = \sqrt{V + 3 m_{3/2}^2 } \, , 
\ee
which is consistent with the results of standard supergravity. 
Notice that $f^S$ is in fact proportional to the {\it effective gravitino mass} 
as defined in \cite{Ferrara:2016ntj} 
\be
m_{3/2}^{\text{eff}} =  \sqrt{|m_{3/2}|^2 + \frac13 V }\, . 
\ee
The simplest way to understand why $f^S$ is given by \eqref{FFSS} 
is by constructing a standard supergravity theory coupled to the nilpotent $X$, 
with \emph{no} higher derivative terms, 
which on the other hand will reproduce exactly \eqref{LL11} in the $G=0$ gauge. 
Indeed to find \eqref{LL11} we would require 
\be
\begin{aligned}
K = & X \overline X \, , 
\\
W = & \sqrt{ |f|^2 -3 |W_0|^2 + 3 |W_0 - \overline \zeta |^2 } \, X + W_0 - \overline \zeta \, . 
\end{aligned}
\ee
The supersymmetry breaking scale now would be $\langle F^X \rangle = \sqrt{ |f|^2 -3 |W_0|^2 + 3 |W_0 - \overline \zeta |^2 }$ 
and the gravitino mass is $m_{3/2} = W_0 - \overline \zeta$.  
Now $\langle F^X \rangle$ will match with $f^S$ given in \eqref{FFSS}, 
and therefore will be consistent with the standard supergravity results.

We close this section with a general discussion on the supersymmetry breaking scale. 
As we have seen there are various operators which can contribute independently to the vacuum energy, 
as for example \eqref{LLcc} and also independently to the gravitino mass as for example \eqref{gmass2}.  
Therefore,  a theory with a non-linearly realized supersymmetry will have a vacuum energy $V$ and  
a gravitino mass $m_{3/2}$. 
Then the supersymmetry breaking scale can be identified with $f^S$ given by \eqref{FFSS}. 
The reason as we explained is simple.  
We can construct such an effective theory from a supergravity coupled to $X$ with no higher derivatives, 
but with K\"ahler potential $K=X \overline X$ 
and superpotential $W = f^S X + m_{3/2}$. 
Notice finally that the models with arbitrary gravitino mass and arbitrary vacuum energy will not necessarily  satisfy the 
unitarity bound on the gravitino mass ($V \geq - 3 m_{3/2}^2$), 
and one has to impose this by hand by adjusting the parameters. 
For a discussion see for example \cite{Cribiori:2016qif}.

Let us also note that we have not considered here
 supergravity matter couplings. In the latter case,  integrating over the gravitino, which now couples  to
matter,   supercurrent-supercurrent interaction are expected to appear since the gravitino becomes an auxiliary field in the zero-momentum limit.

\section{Effective theories}

Now we are in position to couple chiral superfields to the supergravity multiplet in such a way that we generate 
in the bosonic sector couplings of the form \eqref{intro1}. 
Even though one can construct these couplings also within a linearly realized supersymmetric theory, 
the Lagrangian which arises will contain various other terms which give rise to ghosts. 
For example in \cite{Cecotti2}, where the ``supersymmetrization'' of 
the Gauss-Bonnet coupled to a scalar was investigated, 
it was shown that physical scalars and auxiliary fields have kinetic mixings which lead to ghost states. 
However, in our construction, the use of constrained superfields helps to avoid the appearance of 
these states because the fields which leads to these mixings are eliminated from the spectrum. Therefore, the low energy effective theory turns out to be meaningful.

\subsection{Chiral superfield coupled to super Gauss--Bonnet}

As we will now see, the supersymmetric version of 
$a_1 (\phi) \, R^2_{GB}$ 
follows naturally from the coupling of chiral superfields to the supersymmetric Gauss--Bonnet combination  \cite{TNi,Cecotti1,Cecotti2}. 
Before coupling to matter let us remind the reader that, 
as it happens for gravity, 
the Lagrangian that contains the pure Gauss--Bonnet combination is a total derivative, 
therefore contributes only to topological properties.  
The pure super Gauss-Bonnet combination is constructed from the chiral superfield 
\be
\label{YGB}
Y_{GB} = {\cal W}^{\alpha \beta \gamma} {\cal W}_{\alpha \beta \gamma} 
- \frac14 (\overline {\cal D}^2 - 8 {\cal R}) ( 2 \, {\cal R} \overline {\cal R} +  B^a  B_a ) . 
\ee
The form of this combination can be traced back to the linearized supergravity approach \cite{Ferrara:1977mv}. 
An interesting discussion on the properties of $Y_{GB}$ can be also found in \cite{Buchbinder:1988tj}. 
The full component field expansion of \eqref{YGB} can be found in \cite{Cecotti1,Cecotti2}, 
and since it is a chiral superfield it has an expansion of the form 
\be
Y_{GB} = A^Y 
+ \sqrt 2 \, \Theta^\alpha \chi_\alpha^Y 
+ \Theta^2 F^Y .  
\ee 
For our discussion it is important to recall  the bosonic parts of the superfield \eqref{YGB}. 
For the lowest component of \eqref{YGB} we have 
\be
A^Y =   -\frac14 \left( 
 \frac19 M R 
 -\frac{2 i }{9} M  {\cal D}^a b_a  
- \frac{10}{27} M b^a b_a  -\frac{2}{27} M^2 \overline M - \frac{4 i}{9} b^a {\cal D}_a M 
\right) , 
\ee
whereas for the highest component we have  
\be
\label{D2GB}
F^Y =  \frac{1}{16} \left( 
R^2_{GB} - \frac{i}{2} R^2_{HP}  + \frac89 {\cal D}^a P_a \right) \, . 
\ee
The terms with auxiliary fields in \eqref{D2GB}  are  included in 
\be
P_a = M {\cal D}_a \overline M + \frac{i}{3} M \overline M b_a 
+ {\cal D}^b \left( b_a b_b \right) 
- 2 i \, b^b B_{ba} 
+ \frac{2i}{3} \,  b_a b^b b_b  
+ \frac{3i}{2} n_{ab} \epsilon^{bcde} b_c {\cal D}_d b_e \, , 
\ee
where 
\be
B_{ba} = -\frac32 \, \overline \s^{\dot \alpha \alpha}_b {\cal D}_\alpha \overline {\cal D}_{\dot \alpha}    B_a | . 
\ee 
The Lagrangian which gives the pure topological contribution is \cite{Cecotti1,Cecotti2} 
\be
\label{SGB}
{\cal L}_{GB} = 16 \int d^2  \Theta \, 2 {\cal E} \, Y_{GB} + c.c. 
\ee 
The component form of \eqref{SGB} can easily be found by using the generic Lagrangian \eqref{helper}. We find that 
 the bosonic sector of \eqref{SGB} is up to total derivatives 
\be
e^{-1} {\cal L}_{GB} = R^2_{GB} - \frac{i}{2} R^2_{HP} + c.c. 
\ee
Notice that by ``supersymmetrizing'' the Gauss-Bonnet combination one gets also the Hirzebruch--Pontryagin combination $R_{HP}^2$.

Now we will couple supergravity to the chiral superfields $\Phi^I$ via the 
term \cite{Cecotti1,Cecotti2} 
\be
\label{sZGB}
{\cal L}_{\Phi,GB} = 16 \int d^2  \Theta \, 2 {\cal E} \, f(\Phi^I)  \, Y_{GB} + c.c.
\ee
where $f(\Phi^I)$ is a holomorphic function. 
Of course the Lagrangian (\ref{sZGB})  is a total derivative if we set $f(\Phi^I)= {\rm constant}$. 
We use the notation 
\be
f_I = \frac{\p f}{\p A^I} . 
\ee
One can use \eqref{helper} to verify that  
the pure bosonic sector of (\ref{sZGB}) turns out to be
\be
\label{ZGB}
\begin{split}
e^{-1}  {\cal L}_{\Phi,GB} = & \frac89 f_I F_I  \left(  - \frac12 M R + i  M  {\cal D}^a b_a  
+ \frac{1}{3} M b^a b_a +\frac{1}{3} M^2 \overline M \right)  + c.c. 
\\
&+ \frac89 f_I F_I  \left( \frac{4}{3} M b^a b_a + 4 i b^a {\cal D}_a M \right) 
+ c.c. 
\\
& +  f(A^I) \left( R^2_{GB} - \frac{i}{2} R^2_{HP} +  \frac89 {\cal D}^a P_a \right) + c.c. 
\end{split}
\ee

We will employ now the constrained superfields $X$ and ${\cal A}$ introduced in  \eqref{X} and \eqref{AA}, which satisfy 
\be
\label{XXG0}
X^2=0 \ , \ X \left( {\cal A} - \overline {\cal A} \right) = 0 . 
\ee
In addition, we will also use the constrained supergravity multiplet which satisfies both \eqref{XXG} and \eqref{XRc} 
\be
\label{XXG1}  
X \overline X \,  B_a = 0 , ~~~X \left( {\cal R} + \frac{c}{6} \right) = 0 .
\ee 
In the unitary gauge (${G}_\alpha=0$) we have for the matter superfields 
\be
{\cal A} = \phi \  , \  X = \Theta^2 F^X \, , 
\ee
while for the supergravity auxiliary component fields we have 
\be
M = c \ , \ b_a = 0 \ , 
\ee 
where we set $c \in \mathbb{R}$ to simplify the effective theory.
Let us recall that in general the scalar kinetic terms have three 
possible sources: the scalar kinetic term in the original supergavity Lagrangian, the Weyl rescaling, 
and the elimination of the vector $b_a$ of the supergravity multiplet. 
It is the latter integration, which leads to ghost multiplets in the super Gauss--Bonnet supergravity. 
Indeed, as it has been shown in \cite{Cecotti1}, it is this integration that mixes the scalar kinetic terms, 
which after diagonalization leads to ghosts. 
Here, 
by employing the constraint (\ref{XXG1}), which sets $b_a=0$ we remove this mixing and therefore there are no ghosts in this constrained 
super Gauss-Bonnet supergravity. Let us also note that the constraint (\ref{XXG1}) also sets $M=c$ and this simply modifies the scalar potential. 
Indeed, the constraints 
which fix the auxiliary fields of the supergravity multiplet to be $M=c,~b_a=0$, 
modify the standard form of the supergravity. 
For example, in standard supergarvity, before the Weyl rescaling, the Lagrangian with $X\overline X B_a=0$ and $X({\cal R}+c/6)=0$, following \cite{CFGvP} is given by 
\begin{eqnarray}
e^{-1}{\cal L}_{\rm bos}= \frac{1}{6} \Omega \, R 
- \Omega_{I \overline J} \partial_m A^I \partial^m \overline A^{\overline J} 
- V, 
\end{eqnarray}
where 
\begin{eqnarray}
V=\Omega^{I\overline J}\left(\frac{1}{2}W_{\overline J}
+\frac{1}{3}c \, \Omega_{\overline J}\right)\left(\frac{1}{2}W_{ I}+\frac{1}{3}c^* \, \Omega_{I}\right)
-\frac{1}{9}|c|^2 \, \Omega-\frac{1}{2} c W-\frac{1}{2}c^* W.
\end{eqnarray}
After the Weyl rescaling we have 
\begin{eqnarray}
\frac{1}{6}\Omega R \to -\frac{1}{2}R 
-\frac{3}{4}(\partial_m \log \Omega)^2, ~~~
V \to \left(\frac{9}{\Omega^2}\right) V, 
\end{eqnarray}
and the relevant bosonic part of the Lagrangian turns out to be
\begin{align}
e^{-1}{\cal L}_{{\rm bos}(b_a=0,M=c)} = & - \frac12 R 
-\frac{3}{4\Omega^2}\Big(\Omega_I\partial_m A^I+\Omega_{\overline J}\partial_m \overline A ^{\overline J}\Big)^2 
+ \frac{3}{\Omega} \Omega_{I \overline J} \partial_m A^I \partial^m \overline A^{\overline J} 
\\
&
- \left(\frac{9}{\Omega^2}\right) 
\left\{  \Omega^{I\overline J}\left(\frac{1}{2}W_{\overline J}
+\frac{1}{3}c \, \Omega_{\overline J}\right)\left(\frac{1}{2}W_{ I}+\frac{1}{3}c^* \, \Omega_{I}\right)
\right.\nonumber \\
&\left.-\frac{1}{9}|c|^2 \, \Omega-\frac{1}{2} c W-\frac{1}{2}c^* W \right\}.
\label{lagrscalarpot1} 
\end{align}

As a consequence, we see that the metric is not anymore K\"ahler 
(due to the $b_a=0$) and the scalar potential has not the standard
 form of ${\cal N}=1$ supergravity (due to $M=c$).

For a minimal setup, we may assume
\be
\label{KXA}
K = X \overline X - \frac14 ({\cal A} - \overline{\cal A} )^2 , 
\ee
and 
\be
\label{fhg}
W = X \, y({\cal A}) + g({\cal A}) \ , \ f=f({\cal A}). 
\ee 
The functions $y$ and $g$ are both real in the sense that: $y(z)^* = y(\overline z)$. 
The full bosonic sector of the model  
\be
{\cal L} = {\cal L}_0 + {\cal L}_{{\cal A},GB} \, , 
\ee 
turns out to be (after we integrate out $F^X$) 
\be
\label{eff1}
e^{-1} {\cal L} = - \frac12 R 
- \frac12 \p^m \phi \, \p_m \phi  
+ 2 \text{Re}f(\phi) \, R^2_{GB} 
+ \text{Im}f(\phi) \, R^2_{HP} 
-V(\phi) . 
\ee 
The scalar potential has  now a new form 
\be
\label{Vphi}
V(\phi) = \frac13 c^2 + 2 c \, g(\phi) + y(\phi)^2 , 
\ee
due to the constraints we imposed on the gravitational multiplet \cite{Cribiori:2016qif}. This form of the potential coincides with 
Eq.(\ref{sf2}) for real $c,~g(\phi)=m_{3/2},~y(\phi)=f$ of course. 
Notice that by integrating out $F^X$ we have 
\be
F^X = -  y(\phi) 
\ee
therefore, one may use various forms for $y(\phi)$, 
but condition \eqref{condf} must hold. 
We stress also that if we had not 
used the constrained superfield ${\cal A}$, 
which does not have an auxiliary field due to the constraint $|X|^2 {\cal A}=0$, 
the scalar potential would have the form (\ref{lagrscalarpot1}). 
This difference in the structure of the scalar potential has been discussed for example in \cite{calAsuperfield}.

Before concluding this subsection let us discuss the properties of the resultant effective theory \eqref{eff1}. 
Let us recall that  the  independent  quadratic curvature combinations 
are $C_{mnpq}C^{mnpq}$, $R_{mn}R^{mn}$ and $R^2$, where $C_{mnpq}$ is the Weyl tensor. 
From these one can construct the topological 
\begin{eqnarray}
   R^2_{GB}&=& R^{klmn} R_{klmn} -4 R^{mn} R_{mn} + R^2, \label{GB}\\
   R^2_{HP}&=&\epsilon^{klmn}R_{mnqp}{R_{kl}}^{pq}, \label{HP}
   \end{eqnarray}   
namely the Gauss--Bonnet and the Hirzebruch--Pontryagin tensors, which can also be written alternatively  in terms of the Weyl tensor $C_{mnpq}$ as 
\begin{eqnarray}
 R^2_{GB}&=&C_{klmn}C^{klmn} - 2R_{kl}R^{kl} + {2\over 3}R^2, \nonumber \\
   R^2_{HP}&=&\epsilon^{klmn}C_{mnpq}{C_{kl}}^{pq}.   \label{ident1}
\end{eqnarray} 
Both the Gauss--Bonnet and the Hirzebruch--Pontryagin tensors when added in the gravitational action do not contribute to the field equations, which is consistent with the fact that both these tensors are total derivatives  in four-dimensions. 
The contribution of the Gauss-Bonnet and Hirzebruch--Pontryagin tensors is  not trivial once a scalar is coupled to them.

One may consider now the Lagrangian \eqref{eff1},  
where $\text{Re}f(\phi)$ and $\text{Im}f(\phi)$ are arbitrary general functions of (at least) a scalar field $\phi$. 
In this case, there is a contribution to the Einstein  equations, which turns out to be
\begin{eqnarray}
R_{mn}-\frac{1}{2}g_{mn}R=T_{mn}^{\phi}+K_{mn}+C_{mn},
\end{eqnarray}
where 
\begin{eqnarray}
K_{mn}&=&
2R\nabla_m\nabla_n (2 \text{Re}f) 
-2g_{mn}R\nabla^2(2 \text{Re}f) 
\nonumber 
\\
&& -4 R_{mk}\nabla^k\nabla_n(2 \text{Re}f)  
-4 R_{nk}\nabla^k\nabla_m (2 \text{Re}f) 
\nonumber 
\\
&&+4R_{mn}\nabla^2(2 \text{Re}f) 
+4 g_{mn}R_{k\ell}\nabla^k\nabla^\ell (2 \text{Re}f) 
-4 R_{mkn\ell}\nabla^k\nabla^\ell (2 \text{Re}f) , \label{kmm} 
\\
\nonumber
C_{mn}&=&-8\nabla_k (\text{Im}f) \left(e^{pmij}\nabla_iR_j^n+e^{pnij}\nabla_iR_j^m\right)
\\
\nonumber 
&& -4\nabla_p\nabla_q (\text{Im}f) \left( \epsilon^{pnij}{R^{qm}}_{ij}+ \epsilon^{pmij}{R^{qn}}_{ij}\right),
\end{eqnarray}
are the contribution of the variation of the Gauss--Bonnet and Hirzebruch--Pontryagin tensors, respectively and $T^\phi_{mn}$ is the energy-momentum tensor of the scalar.  It is clear from the structure of   $K_{mn}$ that it doesn't contribute with more that two time derivatives to the Einstein equations. Indeed, there are no  terms in (\ref{kmm}) that have more than two derivatives. 
However, the tensor $C_{mn}$ has  a potentially dangerous term which emerges from the derivatives of the Ricci tensor. In particular, for a spacelike $\nabla_k(\text{Im}f)$, one may easily verify that $C_{mn}$ has third time derivatives and therefore, it is potentially problematic. 
However, for timelike $\nabla_k(\text{Im}f)$, there is no such problem as $C_{mn}$ contains only two-time derivatives in this case. For example, on cosmological backgrounds $C_{mn}$ contributes only with second time derivatives. 
In particular, if we have the coupling 
with a scalar profile of the form $
\phi = \phi(t)$,
then the theory will not develop instabilities around a Minkowski background
\footnote{For  theories with stable vacua, as it is the case we are discussing here  ($\phi=\text{const.}$ in the vacuum),  $R^2_{HP}$ does not lead to any linear instabilities and the theory propagates still two polarizations for the graviton. For example, on a Minkowski vacuum with 
background metric $\overline g_{mn}=\eta_{mn}, ~\overline \phi=\phi_0={\rm const.}$, the $R^2_{HP}$ does not contribute to the propagator but it just introduces extra graviton-scalar interactions. The situation is not the same with Lorentz violating backgrounds. The latter have non-vanishing $a_m=\partial_m \phi$ and for spacelike $a_m$  instabilities may occur. 
} \cite{cosmoQG}.

Notice that this coupling $\text{Im}f(\phi) \, R^2_{HP} $ will lead to violation of CP conservation \cite{cosmoQG}.  
In a cosmological framework, parity violating terms appear in the effective theory of inflation \cite{Weinberg:2008hq}. Moreover,  chiral gravity may produce parity-violating TB and EB correlations in the CMB \cite{GK}.

Finally, for  a real form-factor function $f({\cal A})$,  we have 
\begin{eqnarray}
f({\cal A})^*=f({\cal A}^*)
\end{eqnarray}
so that $\text{Im}f(\phi)=0$.  Hence, 
the Lagrangian (\ref{eff1}) 
is written as 
\be
\label{eff2}
e^{-1} {\cal L} = - \frac12 R - \frac12 \p^m \phi \, \p_m \phi  
+ 2 \text{Re}f(\phi) \, R^2_{GB}  
-V(\phi) ,
\ee 
which describes standard gravity with a scalar coupled to the Gauss-Bonnet tensor. This theory propagates normal states and it has been considered as consistent  UV modification of Einstein gravity.

\subsection{Gauss--Bonnet--Volkov--Akulov supergravity}

We now turn to a minimal setup where only the nilpotent superfield $X$ is coupled to supergravity. 
In such a setup the most general coupling is when $X$ has $\Omega$ function and 
superpotential 
\be
\Omega = X \overline X - 3 \ , \  W = f X + W_0 \, ,  
\ee
and for simplicity we assume $f$ and $W_0$ to be real constants. 
For the higher derivative sector the most general form of the coupling to the Gauss--Bonnet is  
\be
\label{sXGB}
{\cal L}_{X,GB} = - 18 \alpha \int d^2  \Theta \, 2 {\cal E} \, X  \, Y_{GB} + c.c. \, , 
\ee
where the constant $\alpha$ can be chosen to be real 
\be 
\alpha=\alpha^* \, . 
\ee
Moreover we want to study an effective theory where for the $X$ superfield we have 
\be
X \overline X \left( - \frac14 \overline{\cal D}^2 \overline X \right) = - f X \overline X \, , 
\ee
which fixes the value of $F^X$ to be (for $G=0$) 
\be
F^X = - f \, . 
\ee
On the supergravity multiplet we impose only \eqref{XXG} which gives (for $G=0$) 
\be
b_a = 0 \, , 
\ee
and we leave the auxiliary field of supergravity $M$ unconstrained. 
Then from Eq.(\ref{ZGB}) we see that the  bosonic sector now reads 
\be
\label{MMM}
\begin{aligned}
e^{-1} {\cal L} = \, &
- \frac12 R - \frac13 |M|^2 - W_0 \overline M - W_0 M 
- f^2 
\\
& 
+ \alpha f \left[ - \frac12 M R -  \frac12 \overline M R + \frac13 M^2 \overline M  + \frac13 \overline M^2 M \right]  \, .   
\end{aligned}
\ee
The first line in (\ref{MMM}) is the bosonic part of the standard supergravity, whereas the second line of (\ref{MMM}) is the modification due to the Gauss--Bonnet term. Notice that this term effects the dynamics  only  when supersymmetry is broken (i.e., $f\neq 0$) 
as was anticipated in \cite{Cecotti2}. 

For the other auxiliary field $M$ we will not  impose \eqref{XRc},  
instead we will  keep it as a dynamical field. In fact, we can integrate it out as it appears algebraically. Before doing that, let us first split
 $M$ into real and imaginary part at 
\be
M = t+is,
\ee
which gives 
\be
\label{uuu}
e^{-1} {\cal L} = 
- \frac12 \left( 1+ 2 \alpha f \, t \right) R - \frac13 t^2 - \frac13 s^2 - 2 W_0 t
- f^2 
+ \frac23 \alpha f t \, (t^2 + s^2 ) \, .   
\ee
Clearly when we integrate out $s$ we find 
\be
s = 0 \, . 
\ee 
After defining the field 
\begin{eqnarray}
t=\frac{1}{2\alpha f}\Big(e^{\gamma \varphi}-1\Big)  \ , ~~~\gamma=\sqrt{2/3}
\end{eqnarray}
and performing a Weyl rescaling of the form 
\be
g_{mn} \rightarrow  e^{-\gamma \varphi} \, g_{mn} \, , 
\ee
we get 
\be
\label{uuu2}
e^{-1} {\cal L} = 
- \frac12 R 
- \frac{ 1}{2} \, \partial_m \varphi \partial^m \varphi 
- V \, ,   
\ee
where the scalar potential is given by
\begin{eqnarray}
V=\frac{1}{12\alpha^2f^2}e^{-2\gamma\varphi}\Big{\{}4(1+3\alpha^2f^4-3\alpha f W_0)+e^{\gamma\varphi}(12\alpha fW_0-11)+10e^{2\gamma\varphi}-3e^{3\gamma\varphi}\Big{\}}. \label{vv}
\end{eqnarray}
A plot of the potential is given below in Fig.1. We see that for large values of $\varphi\to \infty$, $V\sim -e^{\gamma\varphi}$, and for $\varphi\to -\infty$, $V\sim e^{-2\gamma\varphi}$. Moreover, the potential $V$ has  an inflection point.
\begin{figure}[hh]
\begin{center}
\includegraphics[scale=.7]{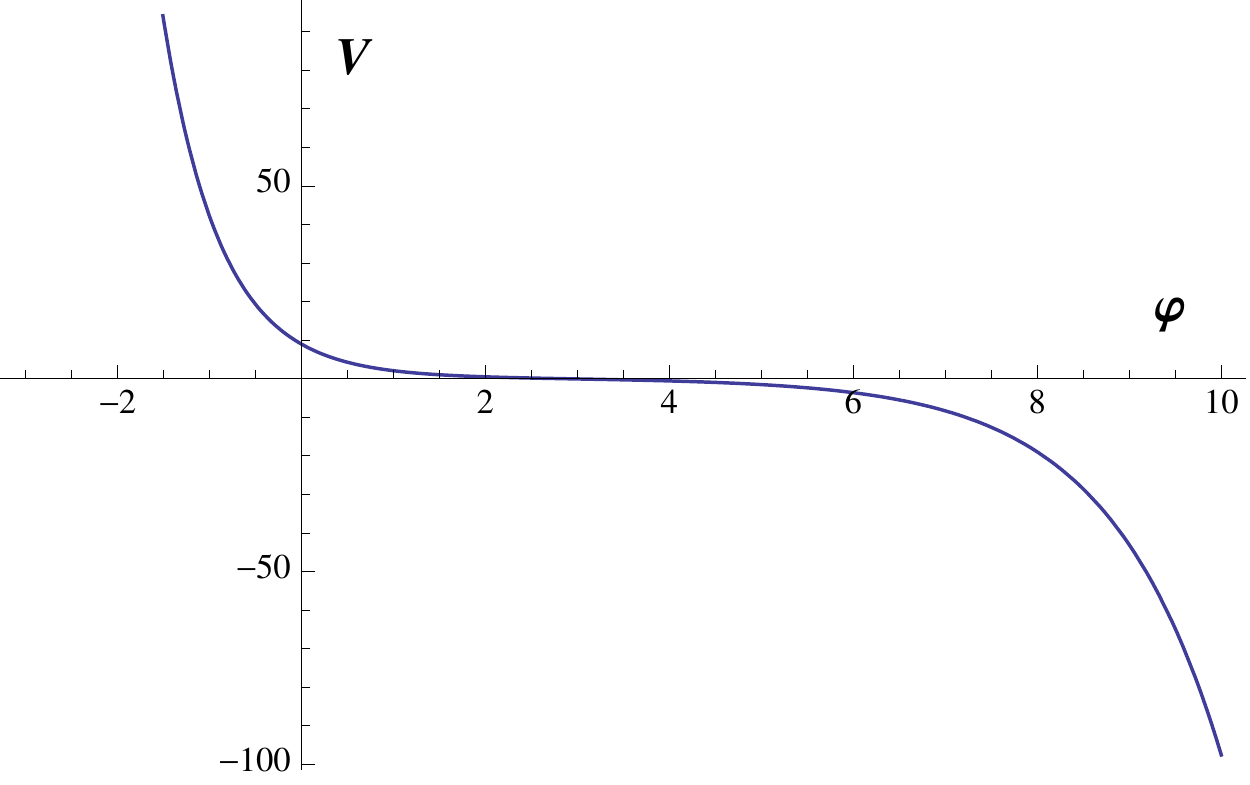}
\caption{The potential $V$ in  (\ref{vv}) for typical values 
of the parameters. Here for
$f=3, \\
\alpha=1,
W_0=4$. }
\end{center}
\end{figure}

\subsection{Chiral superfield coupled to super Weyl square}

In this section we study the supersymmetrization of the square of the Weyl tensor coupled to a scalar, 
by combining the square of the Weyl superfield ${\cal W}_{\alpha\beta \gamma}$ 
with a function of the chiral superfields $\Phi^I$ of the chiral model. 
The Weyl multiplet is chiral therefore we can construct the chiral superfield
\be
{\cal W}^{\alpha \beta \gamma} {\cal W}_{\alpha \beta \gamma} =
A^{{\cal W}} +
\sqrt 2 \, \Theta \chi^{{\cal W}}
+ \Theta^2 F^{{\cal W}} .
\ee
The only contribution to the pure bosonic sector comes from
\be
F^{{\cal W}} = \frac{1}{16} \left( C^{abcd} C_{abcd}
- \frac{i}{2} \, C_{abcd} \epsilon^{cdgh} C_{gh}^{\ \ ab}
- \frac43 F^{ab} F_{ab}
+ \frac{2i}{3} F_{ab} F_{cd} \epsilon^{abcd}
\right) \, ,
\ee
where
\be
F_{ab} = {\cal D}_a b_b - {\cal D}_b b_a \, ,
\ee
and $C_{abcd}$ 
is the Weyl tensor. 
Note that 
$R^2_{HP} = C_{abcd} \epsilon^{cdgh} C_{gh}^{\ \ ab}$. 
The coupling of the chiral superfield to the super-curvature is
described by the term
\be
\label{fw1}
{\cal L}_{\Phi,{\cal W}} = 8 \int d^2 \Theta \, 2{\cal E}
\, h(\Phi^I) \, {\cal W}^2
+c.c.
\ee
From \eqref{helper} we find that the bosonic sector of \eqref{fw1} is
\be
\label{fw2}
e^{-1} {\cal L}_{\Phi,{\cal W}} =  
\text{Im}\, h(A^I) 
\left( \frac12 R^2_{HP}
-\frac{2}{3} F_{ab} F_{cd} \epsilon^{abcd}
\right) 
+\text{Re} h(A^I)   \left( C^{abcd} C_{abcd}
- \frac43 F^{ab} F_{ab}
\right) 
\, .
\ee
In \eqref{fw2} we see that the square of the Weyl tensor is coupled to the real part of the holomorphic function $h$, 
whereas the imaginary part of $h$ is coupled to the Hirzebruch--Pontryagin combination. 
In addition the vector $b_a$ has now become dynamical due to the presence of the kinetic term $- \frac43 F^{ab} F_{ab}$. 
To find the pure quadratic curvature couplings resulting from this coupling we will follow the same procedure as before.

Let us now consider the chiral model \eqref{chiralmodel} with  
K\"ahler potential \eqref{KXA} and superpotential \eqref{fhg}, 
where $X$ satisfies \eqref{X2} and $\cal A$ satisfies \eqref{Ac}. 
Within this setup, 
the field strength of the auxiliary $b_a$ still appears in \eqref{fw2} which we can eliminate 
by removing the field $b_a$ with the constraint \eqref{XXG}. 
In addition, we also impose the constraint \eqref{XRc} on $\cal R$  which will lead us to the same effective 
Lagrangian for the ${\cal L}_0$ sector as in the previous section 
with K\"ahler potential and superpotential given by \eqref{KXA} and \eqref{fhg}. 
Finally we set 
\be
h\left( \Phi^I \right) = h\left( {\cal A} \right) \, , 
\ee
namely that the $h$ holomorphic function depends only on ${\cal A}$. 
Once we add \eqref{fw2} to ${\cal L}_0$ and take into account the constraints,  
the effective theory takes the form 
\be
\label{HPscalar}
e^{-1} {\cal L} = - \frac12 R - \frac12 \p^m \phi \, \p_m \phi
+\frac12 \text{Im} h(\phi) \, R^2_{HP} 
+\text{Re} h(\phi)  C^{abcd} C_{abcd} 
-V(\phi) \, ,
\ee
where $V(\phi)$ is given by \eqref{Vphi}. 
Further properties of the theories with the bosonic sector \eqref{HPscalar} can be found in \cite{cosmoQG}. 
Notice that we did not have to impose the constraints \eqref{XXG} and \eqref{XRc} to eliminate any ghosts, 
we only impose them such that 
the theory becomes simpler and the ${\cal L}_0$ sector takes the form it has in the previous subsection 3.1.

Before concluding this subsection, let us comment on  the cosmological properties of the resulting effective theory \eqref{HPscalar} 
during inflation. 
Assuming that $h({\cal A}) $ is an imaginary function 
\begin{eqnarray}
h({\cal A})^*=-h({\cal A}^*),
\end{eqnarray}
we have that $\text{Re} h(\phi)=0$ and the action  (\ref{HPscalar}) is written as 
\be
\label{HPscalar1}
e^{-1} {\cal L} = - \frac12 R - \frac12 \p^m \phi \, \p_m \phi
+\frac12 \text{Im} h(\phi) \, R^2_{HP} 
-V(\phi) \, . 
\ee
This is the action for chiral gravity which has been discussed in \cite{cosmoQG} and in a cosmological setup in  \cite{Weinberg:2008hq, GK}.
In the opposite case in which $\text{Re} h(\phi)=0$, we are left with 
\be
\label{HPscalar2}
e^{-1} {\cal L} = - \frac12 R - \frac12 \p^m \phi \, \p_m \phi 
- \text{Re} h(\phi)  C^{abcd} C_{abcd} 
-V(\phi) \, .
\ee
This theory suffers from a massive ghost. However, it can be a meaningful theory, if the Weyl term is treated as a perturbation 
to the Einstein--Hilbert term \cite{Weinberg:2008hq}. In this case,
as has been shown in \cite{BPim}, 
the coupling proportional to  $\text{Re} h(\phi)$ will induce a contribution to the tensor tilt 
during slow roll of the form: $n_t \simeq -2 \epsilon \pm 4 \sqrt{2 \epsilon} H^2 \beta$, 
where $\beta=\partial \text{Im} h(\phi) / \partial \phi$. 
This relation will violate the consistency condition $n_t = - r/8$, 
allowing a blue tilt to the tensor spectrum.

\section{Concluding remarks}

In this work we have considered non-linear realizations of local supersymmetry applied to different sectors of a spontaneously broken supergravity  theory. A new constraint removes the gravitino from the spectrum which corresponds to a very massive gravitino in the effective Lagrangian. 
This is the analog of decoupling massive vector states from the low energy spontaneously broken effective gauge theory. Then we consider 
some aspects of non-linear realization in the context of higher curvature supergravity, and in particular of the Gauss--Bonnet  
and Weyl square invariants. When coupled to a chiral field these invariants give rise to ghosts and/or instabilities 
for two different reasons.  The super Gauss--Bonnet introduces ghost degrees of freedom in the sector in which supersymmetry is broken, while the 
Weyl square unavoidably introduces a supermultiplet of a massive spin-2 ghost \cite{Stelle,FGN}.   
For chiral multiplets with the suitable non-holomorphic constraint such instabilities can be avoided and only the  coupling  to the Hirzebruch--Pontryagin  invariant remains.
The other super invariant, the  chiral scalar curvature ${\cal R}^2$ does not suffer from any instabilities and in fact it will propagate two extra physical chiral multiplets.  
An interesting exception is the Volkov--Akulov superfield directly coupled to Gauss--Bonnet. 
In this case, there is an effect due to the appearance of a physical new scalar state coming from the supersymmetric completion of Gauss--Bonnet. 
This mode corresponds to the usual scalar degree of freedom  of a particular  $f(R)$ type theory 
whose dual scalar potential has no extrema.  
This investigation may have some potential implications for cosmological models based on non-linear representations of  local supersymmetry.

\section*{Acknowledgments} 

We thank M. Porrati for enlightening discussions. 
FF was partially supported by the Padua University Project CPDA144437. 
SF was supported in part by the CERN TH-Department and by INFN (IS CSN4-GSS-PI). AK was supported in part by the CERN TH-Department. The work of DL  is supported by the
ERC Advanced Grant No.~320045 ``Strings and Gravity" and by the DFG cluster of excellence "Origin and Structure of the Universe".


\end{document}